\documentclass[preprintnumbers,aps,prl,showkeys,twocolumn,floatfix,nofootinbib,superscriptaddress,groupedaddress]{revtex4}  % for review and submission
\usepackage[paperwidth=215.9mm,paperheight=279.4mm,centering,hmargin=2cm,vmargin=2.5cm]{geometry}
\usepackage[tbtags]{amsmath}  % AMS math 
\usepackage{tabularx}
\usepackage{amssymb}          % AMS symbols
\usepackage{bm}               % bold math
\usepackage{graphicx}         % PostScript figures
\usepackage[export]{adjustbox}% to align images
\usepackage{hhline,multirow}  % for nicer tables
\usepackage{dcolumn}          % Align table columns on decimal point
\usepackage{slashed}
\usepackage{datetime}
\usepackage{color}
\usepackage[dvipsnames]{xcolor}
\usepackage{slashed}
\usepackage{subfloat}

\usepackage{amscd}            % for extensible arrows (e.g., limits)
\usepackage{epsfig}
\usepackage{dcolumn}          % Align table columns on decimal point
\usepackage[dvipsnames]{xcolor}
\usepackage[normalem]{ulem}
\usepackage{mathtools}
\usepackage{comment}
\usepackage[utf8]{inputenc}

\RequirePackage{color}
\RequirePackage[colorlinks=true
,urlcolor=black%blue
,anchorcolor=black%blue
,citecolor=black%blue
,filecolor=black%blue
,linkcolor=black%blue
,menucolor=black%blue
,pagecolor=black%blue
,linktocpage=black%true
,pdfproducer=medialab
,pdfa=true
]{hyperref}

\newcommand{\timeord}{{\cal T}}

\newcommand{\colav}{\frac{\text{Tr}_c}{N_c}}

\newcommand{\Tr}{{\rm Tr}}
\newcommand{\vect}[1]{\boldsymbol{#1}}

\newcommand{\id}{{\mathbb{I}}}
\newcommand{\rom}{{|\Omega\rangle}}
\newcommand{\lom}{{\langle\Omega|}}

\newcommand{\psibar}{{\overline{\psi}}}
\renewcommand{\i}{{\bf i}}

\newcommand{\nocontentsline}[3]{}
\newcommand{\tocless}[2]{\bgroup\let\addcontentsline=\nocontentsline#1{#2}\egroup}

\allowdisplaybreaks[2]
\begin{document}

\preprint{JLAB-THY-19-2900}

\title{Quark fragmentation as a probe of dynamical mass generation}

\author{Alberto Accardi}
\email{accardi@jlab.org}
\affiliation{Hampton University, Hampton, VA 23668, USA}
\affiliation{Theory Center, Thomas Jefferson National Accelerator Facility, 12000 Jefferson Avenue, Newport News, VA 23606, USA}

\author{Andrea Signori}
\thanks{Electronic address: asignori@jlab.org - \href{https://orcid.org/0000-0001-6640-9659}{ORCID: 0000-0001-6640-9659}} 
\affiliation{Physics Division, Argonne National Laboratory, 9700 S. Cass Avenue, Lemont, IL 60439 USA}
\affiliation{Dipartimento di Fisica, Universit\`a di Pavia, via Bassi 6, I-27100 Pavia, Italy}
\affiliation{INFN Sezione di Pavia, via Bassi 6, I-27100 Pavia, Italy}
\affiliation{Theory Center, Thomas Jefferson National Accelerator Facility, 12000 Jefferson Avenue, Newport News, VA 23606, USA}

\date{\today}

\begin{abstract} 
We address the propagation and hadronization of a struck quark by studying the gauge invariance of the color-averaged cut quark propagator,
and by relating this to the single inclusive quark fragmentation correlator
by means of new sum rules. Using suitable Wilson lines, we provide a gauge-invariant definition for the mass of the color-averaged dressed quark and decompose this into the sum of a current and an interaction-dependent component. The latter, which we argue is an order parameter for dynamical chiral symmetry breaking, also appears in the sum rule for the twist-3 $\tilde E$ fragmentation function, providing a specific experimental way to probe the dynamical generation of mass in Quantum Chromo Dynamics.
\end{abstract}

%\pacs{\AScom{add PACS numbers}}

\maketitle

%%%%%%%%%%%%%%%%%%%%%%%%%%%%%%%%%%%%%%%%%%%%%%%%%%%%%%%%%
\textbf{\textit{Introduction.}}
%%%%%%%%%%%%%%%%%%%%%%%%%%%%%%%%%%%%%%%%%%%%%%%%%%%%%%%%%
One of the crucial properties of the strong force is confinement, namely the fact that partons cannot exist as free particles outside of hadrons. As a consequence, any individual parton struck in a high-energy scattering process and freed from its parent hadron must transform into at least one hadron. During this hadronization
process, the colored and much lighter 
parton interacts with the surrounding matter and the vacuum to produce massive and colorless hadrons. Hadronization is thus tightly connected to the dynamical generation of the mass, the spin, and the size of hadrons, but the
%exact
details of the quark-to-hadron transition are still unknown.
Unraveling hadronization dynamics is not only of fundamental importance to understand the nature of visible matter, but also to tackle hadron tomography studies at current and future facilities, such as the 12 GeV upgrade at Jefferson Lab~\cite{Dudek:2012vr} and a future US-based Electron-Ion Collider~\cite{Accardi:2012qut},  where measuring the transverse momentum of one final state hadron is crucial to provide a handle into the transverse motion of quarks and gluons in the hadron target~\cite{Bacchetta:2017gcc,Scimemi:2017etj,Angeles-Martinez:2015sea,Rogers:2015sqa,Bacchetta:2016ccz,Bacchetta:2015ora,Boer:2016xqr,Shi:2018zqd}.

In this letter, we address the propagation and hadronization of a struck quark by studying the gauge invariance properties of the ``inclusive jet correlator'' defined in Ref.~\cite{Sterman:1986aj,Collins:2007ph,Accardi:2008ne,Accardi:2017pmi}, \textit{i.e.}, the color-averaged cut quark propagator supplemented by suitable Wilson lines, and by relating this by means of sum rules to the fragmentation correlator \cite{Bacchetta:2006tn} utilized in Quantum Chromo Dynamics (QCD) to describe the semi-inclusive transition of a quark into a single hadron.

In particular, through the Dirac decomposition of the jet correlator, we provide a gauge-invariant definition of the jet mass $M_j$ that was previously introduced in Ref.~\cite{Accardi:2017pmi}.
In this letter, we elucidate for the first time its nature and properties, recognizing that $M_j$ can be decomposed into the sum of the current quark mass and of an interaction-dependent ``correlation mass''.
%encoding the physics of a color-screened {\em dressed} quark.
Thus the jet correlator can also be interpreted as a color-averaged propagator for a \textit{dressed quark}.

We find that the jet mass and the correlation mass are experimentally accessible through sum rules for the unpolarized collinear twist-3 $E$ and $\tilde E$ fragmentation functions (FFs). A similar sum rule for the $\tilde D^\perp$ FF also supplies information on the average transverse momentum of the produced hadrons. This letter focuses on those FFs whose sum rules encode in a quantitative 
way the intimate connection between hadronization and the dynamical generation of mass and momentum, providing one with a novel way to quantify the quark dressing process, the dynamical breaking of the chiral symmetry, and the nature of the QCD vacuum.

%%%%%%%%%%%%%%%%%%%%%%%%%%%%%%%%%%%%%%%%%%%%%%%%%%%%%%%%%
\ \\ \textbf{\textit{The inclusive jet correlator.}}
%\ \\ \textbf{\textit{The cut quark propagator.}}
%%%%%%%%%%%%%%%%%%%%%%%%%%%%%%%%%%%%%%%%%%%%%%%%%%%%%%%%%
Let us consider the unintegrated color-averaged gauge-invariant cut propagator for a quark of momentum $k$:
%\begin{widetext}
\begin{align}
\label{e:invariant_quark_correlator}
  \begin{split}
  \Xi_{ij}(k;w)
  & = \text{Disc} \int \frac{d^4 \xi}{(2\pi)^4} e^{\i k \cdot \xi} \\
  & \times \colav
    \lom \timeord\, W_1 \psi_i(\xi) \psibar_j(0) W_2 \rom \ , 
  \end{split}
\end{align}
%\end{widetext}
where $\rom$ is the interacting QCD vacuum, and, for simplicity, we omitted the flavor index of the quark field $\psi$.

Gauge invariance is implemented via the Wilson lines $W_1 \equiv W(\infty,\xi;w)$ and $W_2 \equiv W(0,\infty;w)$, and the time-ordering operator $\timeord$ exchanges, when needed, the $\psi$ and $\overline\psi$ fields as well as the ending point $\xi$ of the Wilson line $W_1$ and the starting point $0$ of $W_2$. The vector $w$ will be used to specify the shape of the path in spacetime.

Thanks to the color trace, we can perform a cyclic permutation of the fields. The operator becomes $\psi_i(\xi)\, \overline{\psi}_j(0)\, W(0, \xi; w)$, where the two Wilson lines are now combined into a single one: $W(0, \xi; w) = W_2\, W_1$. 
%
% omit this since it was related to the single Wilson line W_1,2
%We consider L-shaped Wilson lines with displacements to infinity in the $w$ direction, followed by a displacement to infinity in the plane transverse to two light-like vectors $n_+$ and $n_-$, with $n_+ \cdot n_- =1$. 
%
%In this letter, we chose for simplicity of interpretation a light-like $w=n_+$ vector~\cite{Bacchetta:2006tn}, but our techniques can also be applied to off-lightcone vectors $w$ as considered, for example, in~\cite{Collins:2007ph,Collins:2011zzd}. For sake of brevity, we will henceforth omit any explicit dependence on $w$.
%
We consider $W = W_2\, W_1$ to be a staple-like Wilson line,  and the vector $w$ defines the longitudinal direction.
Let us now introduce the light-like vectors $n_+$ and $n_-$ such that $n_+ \cdot n_- =1$. 
For sake of simplicity, in this treatment we restrict our attention to the case $w=n_+$~\cite{Bacchetta:2006tn}, 
but the techniques discussed in this paper can in principle be applied also to an off-the-light-cone vector $w$~\cite{Collins:2007ph,Collins:2011zzd}. 
%For brevity we will henceforth omit any explicit dependence on $w$. 
%
Accordingly, $W$ is a staple-like Wilson line running along the $n_+$ direction to infinity, then jolting 
in the $n_-$ direction and the plane transverse to $n_\pm$,  
%in the plane transverse to the two light-like vectors $n_+$ and $n_-$ such that $n_+ \cdot n_- =1$,  
and running back from infinity again along the $n_+$ direction:   
\begin{widetext}
\begin{align}
  \label{e:W_np}
  W(0, \xi; n_+) & = \,
      {\cal U}[0^+,0^-,\vect{0_T}; \infty^+,0^-,\vect{0_T}]\ 
      {\cal U}[\infty^+,0^-,\vect{0_T}; \infty^+,\xi^-,\vect{\xi_T}]\ 
      {\cal U}[\infty^+,\xi^-,\vect{\xi_T};\xi^+,\xi^-,\vect{\xi_T}]
      \ .
\end{align}
\end{widetext}
On the RHS we have specified the light-cone coordinates of the starting and ending points and we have left the dependence on $n_+$ implicit.  
In Eq.~\eqref{e:W_np} each ${\cal U}$ represents one of the straight segments of the staple:
\begin{align}
  {\cal U}[a;b] = P \exp \Big( -ig \int_a^b dz^\mu A_\mu(z) \Big) \ ,
\end{align}
where $z$ runs in a straight line from $a$ to $b$, $P$ denotes the path-ordering operator, and the square brackets are used to emphasize the straightness of the path.  
%Later on we will work with the gauge-invariant propagator and the quark-to-single-hadron fragmentation correlator integrated over the suppressed component of the partonic momentum, $k^+$, which implies evaluating $W$ in Eq.~\eqref{e:W_np} at $\xi^- = 0$. 
%On the right hand side of Eq.~\eqref{e:W_np}, 

In this letter, we study the cut propagator $\Xi$ as an object of intrinsic theoretical interest in a formal twist expansion in powers of $1/k^-$. This can be justified in scattering processes with a hard scale, such as the 4-momentum transfer $Q$ in deep inelastic scattering. In such processes, the $n_\pm$ vectors are determined by the initial or final state momenta, and $k^- \propto Q \gg k_T \gg k^+$. Our considerations will be, however, process independent.

By taking the discontinuity (Disc) of the propagator -- or its cut, in diagrammatic terms -- Eq.(1) can be interpreted as a gauge-invariant inclusive quark-to-jet %correlator
amplitude squared, or ``inclusive jet correlator'' in short, in which all the quark hadronization products cross the cut and are on-shell~\cite{Sterman:1986aj,Chen:2006vd,Collins:2007ph,Accardi:2008ne,Accardi:2017pmi}.
The adjective ``inclusive'' is to stress that none of the jet's constituents is actually measured, hence the absence of an axis and radius in Eq.~\eqref{e:invariant_quark_correlator}, contrary to semi-inclusive definitions of jets~\cite{Neill:2016vbi,Neill:2018wtk,Gutierrez-Reyes:2018qez,Kang:2017glf,Kang:2017yde,Kang:2017btw}.

The color averaging of the initial-state quark, implemented as Tr$_c[\dots]/N_c$ in Eq.~\eqref{e:invariant_quark_correlator}, arises naturally in QCD factorization theorems and in the definition of the fragmentation correlator, to which the cut quark propagator $\Xi$ is connected through the sum rules we will prove later.
In our analysis, color-averaging plays a crucial role, technically, as it enables a spectral decomposition of the cut propagator $\Xi$. At the conceptual level, it implements the color neutralization that has to take place in order for all the states propagating through the cut to be on-shell hadrons, as confinement dictates for a physical process. 
Therefore, Eq.~\eqref{e:invariant_quark_correlator} can also be interpreted as a {\em color-averaged} (or {\em color-screened}) version of the gauge-invariant Feynman propagator for a dressed quark with four-momentum $k$.
\\

%%%%%%%%%%%%%%%%%%%%%%%%%%%%%%%%%%%%%%%%%%%%%%%%%%%%%%%%%
\textbf{\textit{Spectral decomposition.}}
%%%%%%%%%%%%%%%%%%%%%%%%%%%%%%%%%%%%%%%%%%%%%%%%%%%%%%%%%
The cut propagator $\Xi$ can be given a spectral representation by rewriting Eq.~\eqref{e:invariant_quark_correlator} as a convolution of a quark bilinear $\widetilde S$ and the Fourier transform $\widetilde W$ of the Wilson line $W$: 
%staple-like Wilson line $W \equiv W_2 W_1$:
\begin{align}
\label{e:convolution_Xi}
\Xi_{ij}(k) = \text{Disc} \int d^4p\, 
\colav\,  \lom \widetilde S_{ij}(p) \widetilde W(k-p) \rom \ ,
\end{align}
where
%The Fourier transforms are defined as follows:
\begin{align}
\label{e:def_tildeS}
\widetilde S_{ij}(p) & = \int \frac{d^4\xi}{(2\pi)^4}\, e^{\i\xi \cdot p}\, \timeord\, \psi_i(\xi) \psibar_j(0)\, , \\
\label{e:def_tildeW}
  \widetilde W(k-p) & = \int \frac{d^4\xi}{(2\pi)^4}\, e^{\i\xi \cdot (k-p)}\,
                      \timeord\, W(0,\xi) \, .
\end{align}
In Eq.~\eqref{e:def_tildeS}, the operator $\timeord$ acts as an ordinary time-ordering operator, while in Eq.~\eqref{e:def_tildeW} it acts only on the endpoints of the path for the Wilson line, orienting this from $0$ to $\xi$ or vice versa. 

The quark operator $\widetilde S$ can furthermore be decomposed in Dirac space assuming invariance under Lorentz and parity transformations:%~\cite{Bjorken:1965zz,Accardi:2019sr}
\begin{align}
\widetilde S_{ij}(p) = \hat s_3(p^2) \slashed{p}_{ij} + \sqrt{p^2} \hat s_1(p^2) \id_{ij} \ ,
\label{e:quark_bilinear_decomp}
\end{align}
where we refer to $\hat{s}_{1,3}$ as ``spectral operators''. In principle, when working in an axial $v \cdot A=0$ gauge, we should also add a structure proportional to $\slashed v$ to the right hand side of Eq.~\eqref{e:quark_bilinear_decomp}~\cite{Yamagishi:1986bj}, where $v$ is the vector used to specify the gauge choice, which, in principle, can be different from the vector $w$ used to construct the Wilson line.   
However, in our explicit calculations we adopt the light-cone gauge, $v=n_+ ( = w)$, and, in this specific case, this additional gauge-fixing term would only contribute at twist-4 level,
%~\cite{Accardi:2019sr},
which is not relevant to the matter discussed in this letter.

%The color structure in Eq.~\eqref{e:quark_bilinear_decomp} is an implicit $\id$ in color space. 
%
We can obtain a connection with the usual K\"allen-Lehman spectral representation of the (gauge-variant) quark propagator~\cite{Bjorken:1965zz,Weinberg:1995mt,Accardi:2008ne} by noticing that the quark's Feynman propagator in momentum space is given by the expectation value of $\widetilde S$ on the interacting vacuum. It is then possible to write 
\begin{align}
\label{e:Feyn_spec_rep}
\nonumber 
  \colav\, \lom \widetilde S(p) \rom  
    & = \int_{-\infty}^{+\infty} \frac{d\mu^2}{(2\pi)^4}
    \frac{\i}{p^2-\mu^2+i\epsilon} \\
  & \times \Big\{ \slashed{p}\, \rho_3(\mu^2) + \sqrt{p^2}\, \rho_1(\mu^2) \Big\} \theta(\mu^2) \ .
\end{align}
Using the operator decomposition for $\widetilde S$ given in Eq.~\eqref{e:quark_bilinear_decomp} and the Cutkosky rule~\cite{Cutkosky:1960sp,Bloch:2015efx}, 
we can then connect the spectral operators $\hat{s}_{1,3}$ to the chiral-odd and -even K\"allen-Lehman spectral functions $\rho_{1,3}$:
\begin{equation}
\label{e:spec_rep_s13}
(2\pi)^{3}\, \text{Disc}\, \colav\, \lom \hat{s}_{1,3}(p^2) \rom =  \rho_{1,3}(p^2)\, \theta(p^2)\, \theta(p^-) \ .
\end{equation}

%%%%%%%%%%%%%%%%%%%%%%%%%%%%%%%%%%%%%%%%%%%%%%%%%%%%%%%%%
\textbf{\textit{The TMD inclusive jet correlator.}}
%\textbf{\textit{The integrated cut quark propagator.}}
%%%%%%%%%%%%%%%%%%%%%%%%%%%%%%%%%%%%%%%%%%%%%%%%%%%%%%%%%
%Let us consider, now, the quark propagator integrated over the plus component of the partonic momentum:
It is useful to consider the cut quark propagator integrated over the plus component of the partonic momentum~\cite{Accardi:2017pmi},
%\begin{widetext}
\begin{align}
\label{e:J_TMDcorr}
  J_{ij}(k^-,\vect{k}_T)
    \equiv \frac{1}{2} \int dk^+\, \Xi_{ij}(k) \, ,
%   = \frac{1}{2}\ \text{Disc} \int \frac{d\xi^+ d^2 \xi_T}{(2\pi)^3}\,
%     e^{i k \cdot \xi} \colav
%     \lom W_1 \psi_i(\xi) \psibar_j(0) W_2
%     \rom_{|_{\xi^-=0}} \ ,
\end{align}
where the
% light-cone
time ordering in Eq.~\eqref{e:invariant_quark_correlator} is now trivial since $\xi^-=0$~\cite{Diehl:2003ny,Levelt:1993ac}. 
Likewise, the Wilson line~\eqref{e:W_np} will be evaluated at $\xi^- = 0$. 
As explained later, this integrated correlator is also of interest for the derivation of
% momentum
sum rules for fragmentation functions.

The Transverse Momentum Dependent (TMD) correlator $J$ can be decomposed in Dirac structures, with coefficients determined by Dirac projections of $\Xi$ defined as % given in Eqs.~\eqref{e:trace_gm}-\eqref{e:trace_isigmaipg5}. 
%App.~\ref{a:projections_int} and using Eq.~\eqref{e:proj_int_J}, 
\begin{equation}
\label{e:def_proj}
J^{[\Gamma]} \equiv 
%\Tr\bigg[ J\, \frac{\Gamma}{2} \bigg] = 
\frac{1}{2} \int dk^+ \Tr \bigg[ \Xi\, \frac{\Gamma}{2} \bigg]\, , 
%\frac{1}{4} \int dk^+ \Tr \big[ \Xi\, \Gamma \big] \ .
\end{equation}
where $\Gamma$ is a generic Dirac matrix and Tr represent the trace on the Dirac indexes. 
The structures of interest for the present discussion are:
\begin{align}
\label{e:trace_J_gm}
  \alpha(k^-) & \equiv J^{[\gamma^-]} = \frac{k^-}{\vect{k}_T^i} J^{[\gamma^i]} \ , \\
\label{e:trace_J_id}
  \zeta(k^-) & \equiv \frac{k^-}{\Lambda}  J^{[\id]} \ ,
\end{align}
% \begin{align}
% \label{e:trace_J_gm}
%    J^{[\gamma^-]} & \equiv \alpha(k^-) \ , \\
% \label{e:trace_J_id}
%    J^{[\id]} & \equiv \frac{\Lambda}{k^-} \zeta(k^-)  \ ,
% \end{align}
where $\Lambda$ is a scale with the dimension of a mass introduced for power counting purposes.
%The kinematic dependence of the coefficients and all the remaining Dirac structures will be discussed in detail in Ref.~\cite{Accardi:2019sr}.  
The kinematic dependence of $\alpha$ and $\zeta$ on $k^-$ can be justified as follows. 
The $\Xi(k)$ correlator can be expanded on a basis of Dirac matrices, where the coefficients are, in principle, functions of all the independent scalars that one can build with the available Lorentz vectors, $k$, $n_\pm$~\cite{Boer:2016xqr,Mulders:1995dh}, for example $k \cdot n_+$ and $k^2$. 
Calculating $J^{[\gamma^-]}$ corresponds to integrating over $k^2$ the coefficient of $\Xi$ associated to the $\slashed{k}$ Dirac structure. 
Since this coefficient is only a function of $k^-$ and $k^2$, the result is a simple $k^-$ dependence for $\alpha$. 
A similar argument applies to the definition of $\zeta$. 
%\sout{In the Appendix we also prove the same result by directly evaluating the definition~\eqref{e:trace_J_gm} using Eq.~\eqref{e:convolution_Xi}.}
%The same conclusion can be obtained by calculating $\alpha$ and $\zeta$ directly from tracing Eq.~\eqref{e:convolution_Xi}, as shown in the Appendix. 
%We stress that this is not a general feature: the twist-4 $\gamma^+$ projection, which goes beyond the scope of this letter, is characterized by a coefficient with both a $k^-$ and $k_T$ dependence, since the transverse Wilson line cannot be integrated out. 
%Add physical argument here (no hadron transv. mom. dep. in TMDs!)

%\AScom{Ask Alberto if it's ok. Shall I mention that in \eqref{e:trace_J_gm} I consider only the first definition?} 

%\AScom{I have switched the order of $\alpha$ and $\zeta$ since I wanted to introduce an illustrative appendix for the calculation of $\alpha$}. 
The integrated cut quark propagator
can then be decomposed up to terms of ${\cal O}(\Lambda^2/(k^-)^2)$ as:
\begin{align}
\label{e:J_Dirac}
J(k^-,\vect{k}_T) & = 
\frac{1}{2} \alpha(k^-) \Big[ \gamma^+ + \frac{\slashed{\vect{k}}_T}{k^-} \Big] 
+ \frac{\Lambda}{2k^-} \zeta(k^-) \id \, .
\end{align}

The coefficient $\alpha$ can be calculated directly from the definition~\eqref{e:trace_J_gm}, using the convolution~\eqref{e:convolution_Xi} and the spectral decomposition~\eqref{e:spec_rep_s13}. The details are collected in the Appendix. 
The transverse Wilson line drops out after integration over $d^2\bm{p}_T$ in Eq.~\eqref{e:convolution_Xi}. 
Morever, choosing the light-cone gauge $A \cdot n_+ = 0$ further simplifies the calculation because the longitudinal Wilson lines reduce to the unity matrix in color space. The result is:
\begin{align}
\label{e:alpha_calc_2}
\alpha(k^-) 
& = \frac{\theta(k^-)}{2(2\pi)^3} \bigg\{ \int_0^{+\infty} dp^2 \rho_3^{\,lcg}(p^2) \bigg\} = \frac{\theta(k^-)}{2(2\pi)^3} \ ,
\end{align}
where we have used the gauge-independent normalization property $\int_0^{+\infty} ds \rho_3(s) = 1$ \cite{Weinberg:1995mt} and the superscript $lcg$ stresses the use of the light-cone gauge. 
This confirms that $\alpha$ is a function of $k^-$ only.
%
%Note that one 
%From  we see that one can explicitly factor a 
The $\theta(k^-)$ function factored out in Eq.~\eqref{e:alpha_calc_2} corresponds to 
%$\alpha$ and $\zeta$ because of 
 four-momentum conservation,  
% The positivity of $k^-$ is indeed guaranteed in any gauge by four-momentum conservation,
namely to 
%if one 
assuming that the particles passing the cut all have physical four-momenta.
   
Analogously, we can rewrite the coefficient of the chiral-odd Dirac structure as 
\begin{align}
\label{e:gauge_inv_zeta}
  \zeta(k^-) = \frac{\theta(k^-)}{2(2\pi)^3}\, \frac{M_j(k^-)}{\Lambda} \ ,
\end{align}
where $M_j$ is a {\em gauge-invariant} ``jet'' mass term, potentially function of $k^-$. 
%, and the chosen normalization will become clear after performing the explicit calculation in the light-cone gauge.
As we are going to discuss, this term characterizes the non-perturbative generation of mass in the fragmentation of the quark to the inclusive jet\footnote{The ``jet mass'' $M_j$ was labeled $M_q$ in Ref.~\cite{Accardi:2017pmi} to stress its quark flavor dependence.}.

To illustrate the physical meaning of $M_j$, we calculate $\zeta$ in the light-cone gauge $A \cdot n_+ = 0$. 
%starting from the definition~\eqref{e:trace_J_id}. Using the convolution representation~\eqref{e:convolution_Xi} and the spectral decomposition \eqref{e:spec_rep_s13}, we obtain:
The calculation %of the $\zeta$ coefficient 
follows closely the strategy discussed for $\alpha$ in the Appendix.  
We obtain:
\begin{align}
\label{e:delta_calc_1}
\zeta(k^-) = \frac{\theta(k^-)}{2(2\pi)^3\Lambda} \bigg\{ \int_0^{+\infty} dp^2 \sqrt{p^2} \rho_1^{\,lcg}(p^2) \bigg\} \, ,
\end{align}
which justifies the normalization chosen in Eq.~\eqref{e:gauge_inv_zeta}.
%Comparing Eq.~\eqref{e:delta_calc_1} and~\eqref{e:gauge_inv_zeta}, we can see that
The gauge-invariant $M_j$ mass has a particularly simple and $k^-$-independent form in this  gauge, and is completely determined by the first moment of the spectral function $\rho_1^{\, lcg}$:
\begin{equation}
  \label{e:Mj_lcg}
   M_j  = \int_0^{+\infty} d\mu^2\, \sqrt{\mu^2}\, \rho_1^{\,lcg}(\mu^2) \ .
\end{equation}
The integral at the right hand side
%of Eq.~\eqref{e:Mj_lcg}
is summing over all the discontinuities of the quark propagator. In the light-cone gauge, therefore, $M_j$ can be interpreted as the average mass generated by the chirality flipping component of the quark-to-jet amplitude squared (however, one has to be careful with a probabilistic interpretation because, in a confined theory, the spectral functions are not guaranteed to be positive definite).  
Choosing another gauge, the right hand side of this equation would also receive contributions from the Wilson line, and its physical interpretation would be less immediate. Interpretation aside, we stress that
$M_j$ is gauge-invariant, thanks to the gauge invariance of the operator $\Xi$ in Eq.~\eqref{e:invariant_quark_correlator},
and distinct from the average invariant mass of the fragmented hadrons.
We also note that under renormalization $M_j$ would acquire an additional scale dependence.

Putting all elements together, up to twist 3 we find
\begin{align}
\label{e:J_Dirac}
  4(2\pi)^3 J(k^-,\vect{k}_T)
    & = \Big\{ \gamma^+ + \frac{M_j}{k^-} \id + \frac{\slashed{\vect{k}}_T}{k^-}
      % {\cal O}( k_T/k^- ) + {\cal O}( \Lambda^2/(k^-)^2 )
      \Big\} \, \theta(k^-) \, .
\end{align}
The jet correlator can thus be directly compared to the cut quark propagator with current quark mass $m_{q}$, but now depends on the non-perturbative dressed quark mass $M_j$.

Being related to the trace of the cut propagator, the mass $M_j$ is intrinsically different from the mass function which appears in non-perturbative treatments of the quark propagator~\cite{Roberts:2015lja,Roberts:2007jh}. 
$M_j$ is gauge invariant and scale dependent, whereas the mass function is gauge dependent, but renormalization group invariant. 
Nonetheless, being a scale that characterizes the physics of a color-screened dressed quark, it also provides a window on color confinement.

Remarkably, the gauge-invariant mass $M_j$ is also experimentally accessible because it contributes explicitly to the mass sum rules for the twist-3 fragmentation functions $E$ and $\tilde E$ that we shall prove in the following.

%%%%%%%%%%%%%%%%%%%%%%%%%%%%%%%%%%%%%%%%%%%%%%%%%%%%%%%%%
\ \\ \textbf{\textit{Sum rules for single-hadron fragmentation functions.}}
%\ \\ \textbf{\textit{Momentum sum rules for single-hadron fragmentation functions.}}
%%%%%%%%%%%%%%%%%%%%%%%%%%%%%%%%%%%%%%%%%%%%%%%%%%%%%%%%%
The unintegrated correlator describing the fragmentation of a quark into a single unpolarized hadron is~\cite{Diehl:2003ny,Mulders:1995dh,Goeke:2003az,Bacchetta:2006tn,Meissner:2007rx,Metz:2016swz}:
%\begin{widetext}
\begin{align}
\label{e:1hDelta_corr}
  & \Delta^h_{ij}(k,P_h) = \sum_{S_h} 
  \int \frac{d^4 \xi}{(2\pi)^4} e^{i k \cdot \xi} \\
\nonumber
  & \ \times \colav 
    \lom \timeord W_1 \psi_i(\xi) 
    a_h^\dagger(P_h S_h) a_h(P_h S_h) 
    \overline{\psi}_j(0) W_2 \rom \, ,
\end{align}
%\end{widetext}
where $a_h^\dagger$ and $a_h$ create and destroy a hadron $h$ with momentum $P_h$ and spin $S_h$.

Integrating over the suppressed quark momentum component $k^+$ \cite{Jaffe:1983hp,Diehl:2003ny,Gamberg:2010uw}, one defines the TMD quark-to-single-hadron fragmentation correlator:
%\begin{widetext}
\begin{equation}
\label{e:1hDelta_TMDcorr}
\Delta^h_{ij}(z,\vect{P}_{h\perp}) = 
\int \frac{dk^+}{2z} \text{Disc}\, [\Delta_{ij}^h(k,P_h)]_{k^- = P_h^-/z} \, ,
%\int \frac{d \xi^+ d^2 \xi_T}{2z (2\pi)^3} e^{i k^- \xi^+} 
%\colav\, \text{Disc}\, \lom 
%W_1 \psi_i(\xi) (a_h^\dagger a_h) \overline{\psi}_j(0) W_2  
%\rom_{\begin{subarray}{l} \xi^-=0 \\ k^- = P_h^-/z \end{subarray}}  \ . 
\end{equation}
%\end{widetext}
where, in the parton frame, $\vect{k}_T = 0$ and 
$\vect{P}_{h\perp}$
is the hadronic transverse momentum relative to the quark~\cite{Metz:2016swz,Collins:2011zzd}, 
and $0< z < 1$ if the considered discontinuity is in the $s$-channel of the associated scattering amplitude~\cite{Gamberg:2010uw,Diehl:2003ny}. 
%In the following, all the integrations over $z$ are intended in this range.
%\AScom{In Eq.~\eqref{e:1hDelta_TMDcorr} Disc is the so-called energy discontinuity~\cite{Diehl:2003ny,Gamberg:2008yt}, which implies that $0< z < 1$. In the following, all the integrations over $z$ are intended in this range.}
%\AScom{Maybe mention that the $z$-range is related to the type of discontinuity considered, as explained in~\cite{Gamberg:2010uw}.} 
The parametrization of this correlator in terms of TMD FFs is known up to twist 3~\cite{Mulders:1995dh,Bacchetta:2006tn}. Here we are only interested in
\begin{align}
\label{e:1hDelta_TMDcorr_param}
\Delta^h_{ij}(z,\vect{P}_{h\perp}) & = 
\frac{\gamma^+}{2} D^h_1
+ \frac{M_h}{2 P_h^-} E^h
+ \frac{\vect{\slashed{P}}_{h\perp}}{2z P_h^-} D^{\perp\, h}\, ,
\end{align}
where $D_1^h$ is the unpolarized twist-2 TMD FF and $E^h$ and $D^{\perp\, h}$ are the unpolarized twist-3 TMD FFs.

We can now derive a master four-momentum
sum rule for unintegrated correlators, from which we can later obtain sum rules for specific fragmentation functions.
Let's consider the quantity
\begin{equation}
\label{e:averageh_had_4mom}
%\sum_{h, \, S_h}\, \int \frac{d^4 P_h}{(2\pi)^4}\, (2\pi) \delta(P_h^2 - M_h^2)\, P_h^\mu\, \Delta^h(k,P_h,S_h) \, ,
\sum_{h}\, \int \frac{d^4 P_h}{(2\pi)^4}\, (2\pi) \delta(P_h^2 - M_h^2)\, P_h^\mu\, \Delta^h(k,P_h) \, ,
\end{equation}
which, naively, can be interpreted as the average hadronic four-momentum produced during the hadronization of the parton.  
Introducing the operator associated with the vector $P_h^\mu$~\cite{Meissner:2010cc},
\begin{equation}
\label{e:Ph_op_1}
\hat{\vect{P}}^\mu = \sum_{h, \, S_h} \int \frac{dP_h^- d^2 \vect{P}_{h\perp}}{2P_h^- (2\pi)^3} P_h^\mu \hat{a}_h^\dagger(P_h,S_h) \hat{a}_h(P_h,S_h) \, ,
\end{equation}
and relying on its commutation relation $\big[ {\cal O}(\xi), \hat{\vect{P}}^\mu \big] = \bf{i} \partial^\mu {\cal O}(\xi)$, 
one can relate Eq.~\eqref{e:averageh_had_4mom} to the Fourier transform of the derivative of the matrix element in Eq.~\eqref{e:invariant_quark_correlator}. 
Furthermore, integrating by parts with vanishing boundary conditions for the fields, one obtains:
\begin{equation}
\label{e:master_sum_rule}
\sum_{h} 
\int \frac{d^4 P_h}{(2\pi)^3} \delta(P_h^2 - M_h^2) 
P_h^\mu \Delta^h(k,P_h) = 
k^\mu\, \Xi_{wc}(k) \, , 
\end{equation}
where the subscript $wc$ implies considering $\Xi$ in Eq.~\eqref{e:invariant_quark_correlator} without the discontinuity. 
The relation holds true for all the components $\mu$, but only for $\mu= -,1,2$ there is a connection to the  fragmentation functions. 
%A detailed proof will be presented in Ref.~\cite{Accardi:2019sr}.

We can now obtain sum rules for the fragmentation functions by integrating both sides of Eq.~\eqref{e:master_sum_rule} in the parton frame over the suppressed momentum component $k^+$, 
%which implies calculating 
calculating their discontinuity~\cite{Jaffe:1983hp,Gamberg:2010uw,Diehl:2003ny}, and using suitable Dirac projections of the result. 
Since the sum over the hadron spin $S_h$ is fundamental in the manipulations involving the momentum operator \eqref{e:Ph_op_1}~\cite{Meissner:2010cc}, we can only obtain sum rules for FFs involving unpolarized hadrons.
In particular, for $\Gamma = \{ \gamma^-\, , \id\, , \gamma^i \}$, and suitably choosing $\mu=-$ or $\mu=i$, we obtain:
\begin{align}
\label{e:sumrule_D1}
[\, \Gamma = \gamma^- \, ] \ \ \ \ \ 
& \sum_{h,S_h} \int dz  z\,  D_1^{h}(z) = 1 \ ,	\\
\label{e:sumrule_E}
[\, \Gamma = \id \, ] \ \ \ \ \
& \sum_{h,S_h} \int dz M_h E^{h}(z) = M_j \ , \\
\label{e:sumrule_Dp}
[\, \Gamma = \gamma^i \, ] \ \ \ \ \
& \sum_{h,S_h} \int dz M_h^2 D^{\perp (1) h}(z) = 0 \, .
\end{align}
%The full set of sum rules for FFs up to twist 3 will be presented in Ref.~\cite{Accardi:2019sr}, with preliminary results already discussed at various conferences~\cite{Accardi:2018gmh}.
Note that 
we identified the integrated TMD FFs $f = D_1^h, E^h$ with their collinear counterpart, \textit{i.e.},  $f(z) \ \equiv\ \int d^2\vect{P}_{h\perp} f(z,P_{h\perp}^2)$,
which is formally correct in the context of models of QCD at low energies (given specific regularization prescriptions for the integral over the transverse momentum) and for bare, {\it i.e.} non-renormalized, distributions in perturbative QCD. 
%
%\sout{Note that, 
%since we are working with bare distributions, 
%we identified the integrated TMD FFs $f = D_1^h, E^h$ with their collinear counterpart, \textit{i.e.},  %$f(z) \ \equiv\ \int d^2\vect{P}_{h\perp} f(z,P_{h\perp}^2)$.}
%
%AA , which is correct for the bare distributions.
%While this is correct only when considering bare distributions, the sum rules 
%%\eqref{e:sumrule_D1} and \eqref{e:sumrule_E}
%hold true also in the renormalized case~\cite{Collins:1981uw,Collins:2011zzd}.
% , where in addition one would need to consider the renormalization scale dependence of $M_j$.
We furthermore defined the first moment of any FF $f$ as $f^{(1)}(z) = \int d^2\vect{P}_{h\perp} [\vect{P}_{h\perp}^2/(2z^2M_h^2)]f(z,P_{h\perp}^2)$~\cite{Metz:2016swz}.

%\sout{Despite working with bare quantities, we expect these momentum sum rules to be valid \emph{in form} at the renormalized level, since we based our derivation on the conservation of the partonic four-momentum encoded in Eq.~\eqref{e:master_sum_rule}, and on the symmetry properties of the correlators $\Xi$ and $\Delta$.} 
Since these arguments are related to the conservation of the partonic four-momentum encoded in Eq.~\eqref{e:master_sum_rule} and on the symmetry properties of the correlators $\Xi$ and $\Delta$, we expect these momentum sum rules to be valid \emph{in form} also at the renormalized level in perturbative QCD.    
In fact, renormalization is known to preserve Eq.~\eqref{e:sumrule_D1}~\cite{Collins:1981uw,Collins:2011zzd}, 
% and we postpone a similar study for the other proposed sum rules to a future publication.
and we will address quantitatively the renormalization of the other sum rules in future works.

Eq.~\eqref{e:sumrule_D1} is the well-known momentum sum rule for the unpolarized twist-2 FF, 
and encodes the conservation of the longitudinal momentum in the fragmentation process~\cite{Collins:1981uw}. 
The sum rule~\eqref{e:sumrule_E} was first introduced in Ref.~\cite{Accardi:2017pmi}, but it is proven here for the first time in field theory. It generalizes the sum rule proposed in Ref.~\cite{Jaffe:1993xb}, which however neglected the 
non-perturbative component of the dressed quark mass. We refer to Eq.~\eqref{e:sumrule_E} as the ``mass sum rule'' because of its physical interpretation: the non-perturbative color-screened dressed quark mass $M_j$ corresponds to the average of all the possible masses of the particles produced in the hadronization of the quark, weighted by the chiral-odd collinear twist-3 fragmentation function $E^h(z)$. 
The sum rule \eqref{e:sumrule_Dp} is new, 
and encodes the conservation of the hadronic transverse momentum relative to the momentum of the fragmenting quark.

%%%%%%%%%%%%%%%%%%%%%%%%%%%%%%%%%%%%%%%%%%%%%%%%%%%%%%%%%%%%%%%%%%%%%%%%%%%%%%%%%%%%%%%%%%%%%%%%
\ \\ \textbf{\textit{Twist-three fragmentation and the dynamical component of the jet mass.}}
%%%%%%%%%%%%%%%%%%%%%%%%%%%%%%%%%%%%%%%%%%%%%%%%%%%%%%%%%%%%%%%%%%%%%%%%%%%%%%%%%%%%%%%%%%%%%%%%
Let us now consider the equations of motion relations discussed in Ref.~\cite{Bacchetta:2006tn}, that relate the twist-2 and the twist-3 fragmentation functions. In particular, we need
\begin{align}
\label{e:eom_E}
& E^h = \tilde{E}^h + z \frac{m_{q}}{M_h} D^h_1 \, , \ \ \ \ \ D^{\perp\, h} = \tilde{D}^{\perp\, h} + z D^h_1 ,
\end{align}
where the functions with a tilde are related to the parametrization of the dynamical twist-3 quark-gluon-quark correlator of type $\tilde{\Delta}_A^\alpha$~\cite{Bacchetta:2006tn} and $m_{q}$ is the current quark mass. 

Within the ``Wandzura-Wilczek (WW) approximation''~\cite{Bastami:2018xqd},
namely assuming that the $\tilde E$, $\tilde D^{\perp}$ FFs are negligible compared to the quark-quark FFs without the tilde, $M_j$ reduces to the current quark mass, 
as can be seen by setting $\tilde E = 0$ in Eq.~\eqref{e:eom_E} and using the sum rules~\eqref{e:sumrule_D1} and~\eqref{e:sumrule_E}. This suggests a decomposition of $M_j$ into the sum of the current quark mass and of an interaction-dependent (or dynamical) mass $m_q^{corr}$ generated by quark-gluon-quark correlations, which we name ``correlation mass'':
\begin{equation}
\label{e:Mj_decomp}
M_j = m_{q} + m_q^{corr} \ .
\end{equation}
In analogy with the generation of the quark mass induced by dynamical chiral symmetry breaking~\cite{Roberts:2007jh,Roberts:2015lja}, we expect 
%that the dynamical mass 
$m_q^{corr} = O(\Lambda_{CQD})$. For light quarks, therefore, $M_j$ might be substantially larger than the current quark mass $m_{q}$. 
Combining the sum rules~\eqref{e:sumrule_D1} and~\eqref{e:sumrule_E} with Eq.~\eqref{e:eom_E} and~\eqref{e:Mj_decomp} we obtain:
\begin{align}
\label{e:sumrule_Et}
& \sum_{h\, S_h} \int dz M_h \tilde{E}^{h}(z) = m_q^{corr} \, ,
\end{align}
and one clearly sees that $m_q^{corr}$ is generated by the interaction-dependent quark-gluon-quark correlations encoded 
in the $\tilde E$ function. 
Thus, Eq.~\eqref{e:sumrule_Et} generalizes the sum rule $\int dz \tilde{E} = 0$ discussed in Ref.~\cite{Bacchetta:2006tn} to the non-perturbative case.

The correlation mass vanishes when neglecting quark-gluon-quark correlations, as in the already discussed WW approximation. However, when interactions are considered, the spectral representation \eqref{e:Mj_lcg} shows that $m_q^{corr}$ and $M_j$ receive contributions from all possible propagating hadronic states, including baryons. As a consequence, contrary to $m_q$, we expect that $m_q^{corr}$ and $M_j$ do not vanish even in the chiral limit. 
Therefore, they can also be considered order parameters for dynamical chiral symmetry breaking and are moreover experimentally accessible, at least in principle, via the mass sum rule for the $E$ and $\tilde E$ FFs. 

Finally, combining the sum rules~\eqref{e:sumrule_D1} and~\eqref{e:sumrule_Dp} with the $D$-type equation of motion in Eq.~\eqref{e:eom_E}, we obtain:
\begin{align}
\label{e:sumrule_Dtp}
\nonumber \sum_{h\, S_h} \int dz M_h^2 \tilde{D}^{\perp\, (1)\, h}(z) & = 
- \sum_{h\, S_h} \int dz z\, M_h^2 D_1^{(1)\, h}(z) \\  & 
\equiv \ -\frac{1}{2} \langle \vect{P}_{h\perp}^2 / z \rangle \, .
\end{align}
This new sum rule provides an experimental way of accessing the average squared transverse momentum acquired by unpolarized hadrons fragmenting off an unpolarized quark, and probes the dynamical nature of the hadronization process in analogy with the way the sum rule for $\tilde E$ probes dynamical chiral symmetry breaking. 
Indeed, the transverse momentum acquired during the hadronization is generated by the quark-gluon-quark correlations and would vanish in the absence of these, as it can be directly checked assuming the WW approximation.  
Similar sum rules involving transversely polarized quarks or hadrons have been discussed in Refs.~\cite{Metz:2016swz,Qiu:1991pp}.

%%%%%%%%%%%%%%%%%%%%%%%%%%%%%%%%%%%%%%%%%%%%%%%%%%%%%%%%%
\textbf{\ \\ \textit{Summary and outlook.}}
%%%%%%%%%%%%%%%%%%%%%%%%%%%%%%%%%%%%%%%%%%%%%%%%%%%%%%%%%
%
With the jet mass $M_j$ defined in Eqs.~\eqref{e:trace_J_id} and~\eqref{e:gauge_inv_zeta}, we have for the first time proposed a \textit{gauge-invariant} definition of a color-screened dressed quark mass, which is furthermore experimentally accessible by using the 
proposed 
sum rules~\eqref{e:sumrule_E} and~\eqref{e:sumrule_Et} for twist-3 collinear FFs. In perturbation theory, $M_j$ is proportional to the current quark mass, but in the full theory it remains non-zero also in the chiral limit. Therefore we also recognize $M_j$ as an order parameter for dynamical chiral symmetry breaking.
This provides a novel connection between hadronization and the dynamical generation of mass in QCD.

The dynamical component of the jet mass, quantified by the correlation mass $m_q^{corr}$ in Eq.~\eqref{e:Mj_decomp}, is generated by the quark-gluon-quark correlations encoded in the $\tilde E$ fragmentation function.
% , and is intrinsically connected to the properties of the QCD vacuum.
The $\tilde E$ FF, and thus the sum rule \eqref{e:sumrule_Et} connecting this to $m_q^{corr}$, can be experimentally accessed by looking at chiral odd observables at twist-3 level and higher in semi-inclusive hadron production in polarized electron-hadron scattering, electron-positron annihilation, and hadronic collisions.
As proposed in Ref.~\cite{Accardi:2017pmi,Accardi:2018gmh}, the correlation mass itself may also directly contribute to the inclusive DIS $g_2$ structure function at large Bjorken $x$, and to an analogous asymmetry in electron-positron induced dihadron production.
More generally, in scattering processes, the jet mass $M_j$ (rigorously defined as the trace of the gauge-invariant cut quark propagator) appears to play a similar phenomenological role to that of the constituent mass in quark models. 

With the sum rule~\eqref{e:sumrule_Dtp} 
we are also suggesting that the transverse momentum acquired during the hadronization process is a fully dynamical quantity, namely it is generated by the quark-gluon-quark correlations and it vanishes in the absence of these.
However, like for Eq.~\eqref{e:sumrule_Et}, this interpretation needs to be corroborated by a study of the renormalization properties of the involved operators. 
Such a sum rule could be, in principle, experimentally accessed at twist-2 level through the TMD FF $D_1^h$, about which limited information is so far available~\cite{Matevosyan:2011vj,Bacchetta:2015ora,Bentz:2016rav,Boglione:2017jlh,Moffat:2019pci}. 

Finally, we note that the master sum rule~\eqref{e:master_sum_rule} and the calculational techniques we have introduced have a general applicability. In particular, we expect the convolutional spectral representation of the quark correlator to be also applicable to the study of the gauge invariance of other correlators, for example the recently introduced virtuality-dependent parton distributions~\cite{Radyushkin:2016hsy} that play an important role in the direct lattice QCD calculation of PDFs in momentum space~\cite{Orginos:2017kos,Karpie:2017bzm}. Moreover, our methods are not limited to
%staple-like
Wilson lines on the light-cone, making their domain of applicability potentially wide.

%%%%%%%%%%%%%%%%%%%%%%%%%%%%%%%%%%%%%%%%%%%%%%%%%%%%%%%%%
\textbf{\ \\ \textit{Appendix - calculation of the $\alpha$ coefficient.}}
%%%%%%%%%%%%%%%%%%%%%%%%%%%%%%%%%%%%%%%%%%%%%%%%%%%%%%%%%
\label{app:alpha}
We outline here the steps involved in the calculation of the $\gamma^-$ projection of the correlator $J$ in Eq.~\eqref{e:J_TMDcorr}.  

Using the definition of $\alpha$ given in Eq.~\eqref{e:trace_J_gm} and the convolution representation for $\Xi$ in Eq.~\eqref{e:convolution_Xi} and~\eqref{e:quark_bilinear_decomp}, we find:
\begin{widetext}
\begin{align}
\label{e:alpha_calc_1}
\alpha(k^-) & 
%= \frac{1}{4} \int dk^+ \Tr \big[ \Xi(k) \gamma^- \big] 
= \int dk^+ \text{Disc} \int_M d^4p \colav \lom \hat{s}_3(p^2) p^- \widetilde W(k-p) \rom \\
\nonumber
& = \text{Disc} \int_{\mathbf R} dp^2 \int_{\mathbf R} \frac{dp^-}{2} \colav 
\lom \hat{s}_3(p^2) \int \frac{d\xi^+}{2\pi} e^{i \xi^+ (k^- - p^-)} W_{coll}(\xi^+) \rom \ ,
\end{align}
\end{widetext}
where the integration domain for $p$ is the whole Minkowski space ($M$) and we decompose the integral as $d^4p =  dp^2\, d^2\vect{p}_T\, dp^-/2p^-$. 
From the first to the second line we used Eq.~\eqref{e:def_tildeW} and performed the integrations over $k^+$ and $\vect{p}_T$, fixing $\xi^-=0$ and $\vect{\xi}_T=0$ so that the 
%staple-shaped TMD 
Wilson line reduces to the collinear Wilson line $W_{coll}(\xi^+) \equiv W(0^+,0^-,\bm{0_T}; \xi^+,0^-,\bm{0_T})$. 
Accordingly, we loose any $k_T$ dependence for the $\alpha$ coefficient, and the same argument applies to the calculation of the $\zeta$ coefficient. 
Next, we choose the light-cone gauge $A^-=0$ so that the collinear Wilson line reduces to the unity matrix in color space. Finally, performing the integration over $\xi^+$ we obtain:
% \label{e:alpha_calc_2}
\begin{widetext}
\begin{align}
\alpha(k^-) & = \text{Disc} \int_{\mathbf R} dp^2 \int_{\mathbf R} \frac{dp^-}{2} \colav\, \lom \hat{s}_3(p^2) \rom \delta(k^- - p^-) 
 = \int_{\mathbf R} dp^2 \int_{\mathbf R} \frac{dp^-}{2} \delta(k^- - p^-) \{ (2\pi)^{-3} \rho_3(p^2) \theta(p^2) \theta(p^-) \} \\
\nonumber
& = \frac{1}{2(2\pi)^3} \bigg\{ \int_0^{+\infty} dp^2 \rho_3(p^2) \bigg\} \theta(k^-) = \frac{\theta(k^-)}{2(2\pi)^3} \ ,
\end{align}
\end{widetext}
where we have used the representation for the spectral operator $\hat{s}_3$ given in Eq.~\eqref{e:spec_rep_s13}, and in the last step we used the normalization property for $\int_0^{+\infty} d\mu^2 \rho_3(\mu^2) = 1$. 
We remark that the only dependence on $k^-$ resides in the theta function.  
%which appears due to four momentum conservation. 
%This function and the accompanying coefficients are also a feature of the higher-twist $\zeta$ and $\omega$ coefficientts to be discussed next.

%%%%%%%%%%%%%%%%%%%%%%%%%%%%%%%%%%%%%%%%%%%%%%%%%%%%%%%%% 
\textbf{\ \\ \textit{Acknowledgments.}}
%%%%%%%%%%%%%%%%%%%%%%%%%%%%%%%%%%%%%%%%%%%%%%%%%%%%%%%%%
We thank I. Clo\"et, J. Goity, A. Pilloni, J.W. Qiu, C. Roberts, C. Shi for stimulating discussions. 
This work was supported by the  U.S. Department of Energy contract DE-AC05-06OR23177, under which Jefferson Science Associates LLC manages and operates Jefferson Lab. 
AA also acknowledges support from DOE contract DE-SC0008791.
AS also acknowledges support from the U.S. Department of Energy, Office of Science, Office of Nuclear Physics, contract no. DE-AC02-06CH11357, 
and by the European Commission through the Marie Sk\l{}odowska-Curie Action SQuHadron (grant agreement ID: 795475).  

%%%%%%%%%%%%%%%%%%%%%%%%%%%%%%%%%%%%%%%%%%%%
%\bibliographystyle{JHEP}
\bibliography{JSR}
%%%%%%%%%%%%%%%%%%%%%%%%%%%%%%%%%%%%%%%%%%%%

\end{document}